\documentclass[10pt, paper=a4, UKenglish]{article}
\pdfoutput=1
\usepackage{graphicx}
\def\Title#1{\begin{center} {\Large #1 } \end{center}}
\def\Author#1{\begin{center}{ \sc #1} \end{center}}
\def\Address#1{\begin{center}{ \it #1} \end{center}}

\newcommand\pubblock{\rightline{\begin{tabular}{l} Proceedings of the CTD/WIT 2019\\ \pubnumber\\
         \pubdate  \end{tabular}}}

\newenvironment{Abstract}{\begin{quotation} \begin{center} 
             \large ABSTRACT \end{center}\bigskip 
      \begin{center}\begin{large}}{\end{large}\end{center} \end{quotation}}

\newenvironment{Presented}{\begin{quotation} \begin{center} 
             PRESENTED AT\end{center}\bigskip 
      \begin{center}\begin{large}}{\end{large}\end{center} \end{quotation}}

\def\Acknowledgements{\bigskip  \bigskip \begin{center} \begin{large}
      \bf ACKNOWLEDGEMENTS \end{large}\end{center}}

\def\beq{\begin{equation}}
\def\eeq#1{\label{#1}\end{equation}}
\def\eeqn{\end{equation}}

\def\beqa{\begin{eqnarray}}
\def\eeqa#1{\label{#1}\end{eqnarray}}
\def\eeqan{\end{eqnarray}}

\let\bar=\overbar

\def\Dslash{\not{\hbox{\kern-4pt $D$}}}
\def\dslash{\not{\hbox{\kern-2pt $\del$}}}

\def\msb{{\bar{\ssstyle M \kern -1pt S}}}

\usepackage{color}
\usepackage{lineno}
\usepackage{subfig}
\usepackage{hyperref}
\usepackage{float}

\newcommand\pubnumber{PROC-CTD19-104}

\newcommand\pubdate{\today}

\newcommand{\conference}{Connecting the Dots and Workshop on Intelligent Trackers (CTD/WIT 2019)\\
Instituto de F\'isica Corpuscular (IFIC), Valencia, Spain\\ 
April 2-5, 2019}

\usepackage{fancyhdr}
\definecolor{mygrey}{RGB}{105,105,105}
\begin{document}

% uncomment the following line for adding line numbers
% \linenumbers

% large size for the first page
\large
\begin{titlepage}
\pubblock

%% Change the title, name, abstract
%% Title 
\vfill
\Title{Use of R-trees to improve reconstruction time in pixel trackers}
\vfill

%  if you need to add the support use this, fill the \support definition above. 
%  \Author{FIRSTNAME LASTNAME \support}
\Author{Albert Pern\'\i a V\'azquez}
\Author{N\'uria Valls Canudas}
\Author{Elisabet Golobardes Rib\'e}
\Author{Alessandro Camboni}
\Author{Xavier Vilas\'\i s-Cardona}
\Address{DS4DS, La Salle-Universitat Ramon Llull \\ Quatre Camins 30, 08022 Barcelona, Spain}
\vfill

\begin{Abstract}
    Computing time is becoming a key issue for tracking algorithms both online and off-line. Programming using adequate data structures can largely improve the efficiency of the reconstruction in terms of time response. We propose using one such data structure, called R-tree, that performs a fast, flexible and custom spatial indexing of the hits based on a neighbourhood organisation. The overhead required to prepare the data structure shows to be largely compensated by the efficiency in the search of hits that are candidate to belong to the same track when events present a large number of hits. The study, including different indexing approaches, is performed for a generic pixel tracker largely inspired in the upgrade of the LHCb vertex locator with a backwards reconstruction algorithm of the cellular automaton type.  
\end{Abstract}

\vfill

% DO NOT CHANGE!!!
\begin{Presented}
\conference
\end{Presented}
\vfill
\end{titlepage}
\def\thefootnote{\fnsymbol{footnote}}
\setcounter{footnote}{0}
%

% normal size for the rest
\normalsize 

%% Your paper should be entered below. 

% EO: deixo això per ara com a exemple
%% EEOO : jo ho comento per veure com anem de pagines
%\input{CTDITTemplate/example.tex}

\section{Introduction}
\label{section_intro}

The evolution of silicon technologies in particle detectors has lead to design tracking sensors divided in pixels which allow higher precision. The main LHC experiments have or shall have in the coming future one of such devices \cite{AtlasPixel,CMSPixel,LHCbPixel}. These type of trackers convey much more data, especially at high luminosities, such as the ones we find in the LHC and its high luminosity version HL-LHC. In this context, reconstruction algorithms need to cope with very high data rates in order to timely deliver the required precision. To achieve this requirement, a proper programming and use of adequate data structures play a capital role. 

Our proposal is an elaboration on a well-known reconstruction algorithm, the Cellular Automaton \cite{cellular_automaton}, which performs a backwards pixel search to tracks. This search can be optimised by using a data-structure called R-tree \cite{r_trees_book_theory_and_applications} that performs a fast, flexible and custom spatial indexing of the hits based on a neighbourhood organisation. We show that the overhead required to prepare the data structure is largely compensated by the efficiency in the search of hits that are candidate, to belong to the same track when events present a large number of hits. The study, including different indexing approaches, is performed for a generic pixel tracker largely inspired in the upgrade of the LHCb vertex locator \cite{VeloUpgradeTDR}.

This note is structured as follows. We first give a general overview of track reconstruction in pixel detectors, particularly focused on the cellular automaton algorithm. Then, we describe R-trees and how can they be used to obtain a faster reconstruction, with a general algorithm and with an {\it ad hoc} heuristics. Next, we present the detector model used to test the performance of our proposal, which is followed by results and conclusions. 
\section{Pixel track reconstruction}
\label{section_pixel_track_reco}

A widely used method for track reconstruction in pixel trackers is the so-called Cellular Automaton (CA) algorithm. As the name suggests it is closely inspired in the cellular automaton model used to simulate discrete systems formed by a set of simple objects that interact locally with each other. Moving this concept into track reconstruction, one can treat the data hits as simple objects and check its neighbourhood in order to determine which of them belong to the same track. In other words, the closest hits among different modules, in terms of spatial distance, will be the best candidates to form a segment of a track.

Many adaptations of this method have been implemented, such as in  \cite{event_reconstruction}. In this paper, a much simplified version will be used. More in detail, the flow of the algorithm starts by looking at the two farthest modules from the collision point. By doing this, one can make sure that the hits in them will be as separated as possible in the transverse plane with respect to the $Z$ axis. The first procedure executed in the algorithm is the so-called \textit{seeding}, which consists of finding all the pairs of hits, one in each of the two last modules, that pass a certain criterion of proximity. If a pair does, it will be marked as a \textit{seed} and so become a strong candidate to be the initial part of a forming track.

Once a seed is found, the algorithm starts looking sequentially at the next adjacent modules of the detector in a backward direction, checking for compatibility between every hit and the seed. The compatibility criterion stands for a hit to fit in the line formed by the seed within a certain tolerance based on scattering properties. If a hit is compatible, it is added to the seed to form a triplet and the algorithm keeps searching for compatible hits among the following modules until either the first module of the detector is reached or the algorithm finds three consecutive modules without any compatible hit in them. If any of the previous situations happen, the algorithm concludes to have a reconstructed track and repeats the same procedure for all the seeds found in the last two sensors. All the hits that are already reconstructed as part of a track are marked as used and not checked again.

Finally, once all the seeds are processed, the algorithm repeats the whole procedure but moving one module backwards. This means it will search for seeds in the third and second modules from the back. It keeps on moving one module closer to the front of the detector on each iteration until the fourth module is reached. The full event reconstruction will be finished once all the seeds found in the fourth and fifth module, starting from the front of the detector, are processed.

As previously explained, the seeding procedure matches pairs of hits from adjacent modules based on the distance between them. The main problem of this step is that the only known information about the hits is its position inside the detector, given in three coordinate system. In other words, it only gives information about the local position of each hit, but not about its position related to others. So, in order to find the neighbours of a single hit, one must compute the distance between it and the rest of the hits. Computationally, this is translated into a loop of $N-1$ distance calculations for every hit in a module, where $N$ is the total number of hits in the adjacent module of the hit.

It seems hard to believe that only knowing the position of each hit inside the detector the relation between them is never known. In this paper, we try to add this spatial information to the data using a data structure called R-tree in order to index the hits. This implies a notable reduction of the number of hits evaluated during the candidate search for each hit.  

\section{R-trees}
\label{sec_r_trees}

% Intro
\subsection{The concept}

\begin{figure}[h]
  \centering
  \hspace*{-0.0in}
  \includegraphics[scale=.75]{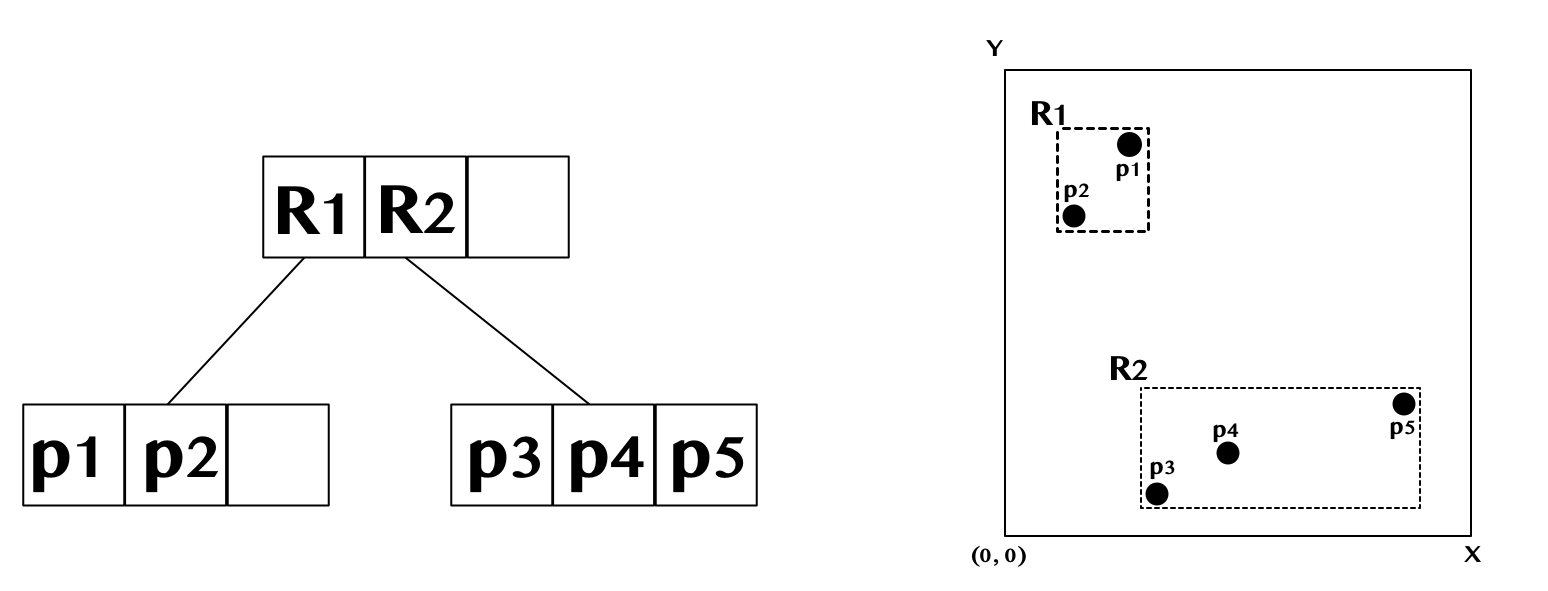}
  \caption{Example of a 2D R-tree with $M = 3$ storing point objects.}
  \label{fig:rtree_example}
\end{figure}

The original R-tree \cite{guttman} represented in figure \ref{fig:rtree_example} consists of a height-balanced tree similar to a B-tree that allows the indexing of \textit{d}-dimensional spatial objects based on their neighbourhood. The idea is to group near objects by MBR (Minimum Bounding Rectangle), being those MBRs represented by bigger ones reiteratedly.

The purpose of this methodology is to be able to retrieve data based on its location without the need to check all the objects stored. Therefore, questions like $"$get all the objects in a specific area$"$ can be answered optimally using such data structure.

An R-tree of order $(m, M)$ has the following properties \cite{r_trees_book_theory_and_applications}:

\begin{itemize}
    \item All the leaf nodes can store between $m$ and $M$ entries unless the leaf node is the root. Each entry is formed by an MBR and a pointer to the associated child node.
    \item All the inner nodes contain between $m$ and $M$ entries unless the inner node is the root. Each entry is formed by an MBR and a pointer or identifier of the object.
    \item The root node must hold at least two entries if it is an inner node. Otherwise, it can be empty.
    \item All the leaves must be at the same level.
\end{itemize}

The maximum height of the R-tree is $log_m(N) - 1$, where $N$ is the total number of objects held by the structure.

\subsection{Insertion}

The insertion procedure of the original R-tree is well described in its original article \cite{guttman}. Although, a brief explanation of the insertion methodology is explained below.

In order to put a spatial data object into the structure, a leaf node is selected to hold it. The procedure starts from the root node and is summarised by the following steps:
\begin{enumerate}
    \item Check whether the current node is a leaf or an inner node. If it is a leaf, then execute step 2, else execute step 3.
    \item Insert the object and propagate the changes upwards. Finish the procedure. 
    \item Else, choose the best entry \textit{e} from the current node to keep the search. The best entry will be the one that its area increment is the minimum needed to cover the new object. In the case of draws, the MBR with the lowest area is chosen. Execute step 4.
    \item Let the current node be $e_{child}$, where $child$ is the pointer to the child node of the chosen entry. Go step 1.
\end{enumerate}

\subsubsection*{Full node case}

If a node becomes full, the new entry cannot be contained. Therefore, a split procedure is executed in order to distribute the existing entries plus the new one in two resulting nodes. These two new nodes will be also covered by two MBRs on the parent node. This explanation is exemplified below in figure \ref{fig:tree_insertion_full_node}.

\begin{figure}[H]
  \centering
  \hspace*{-0.0in}
  \includegraphics[scale=0.62]{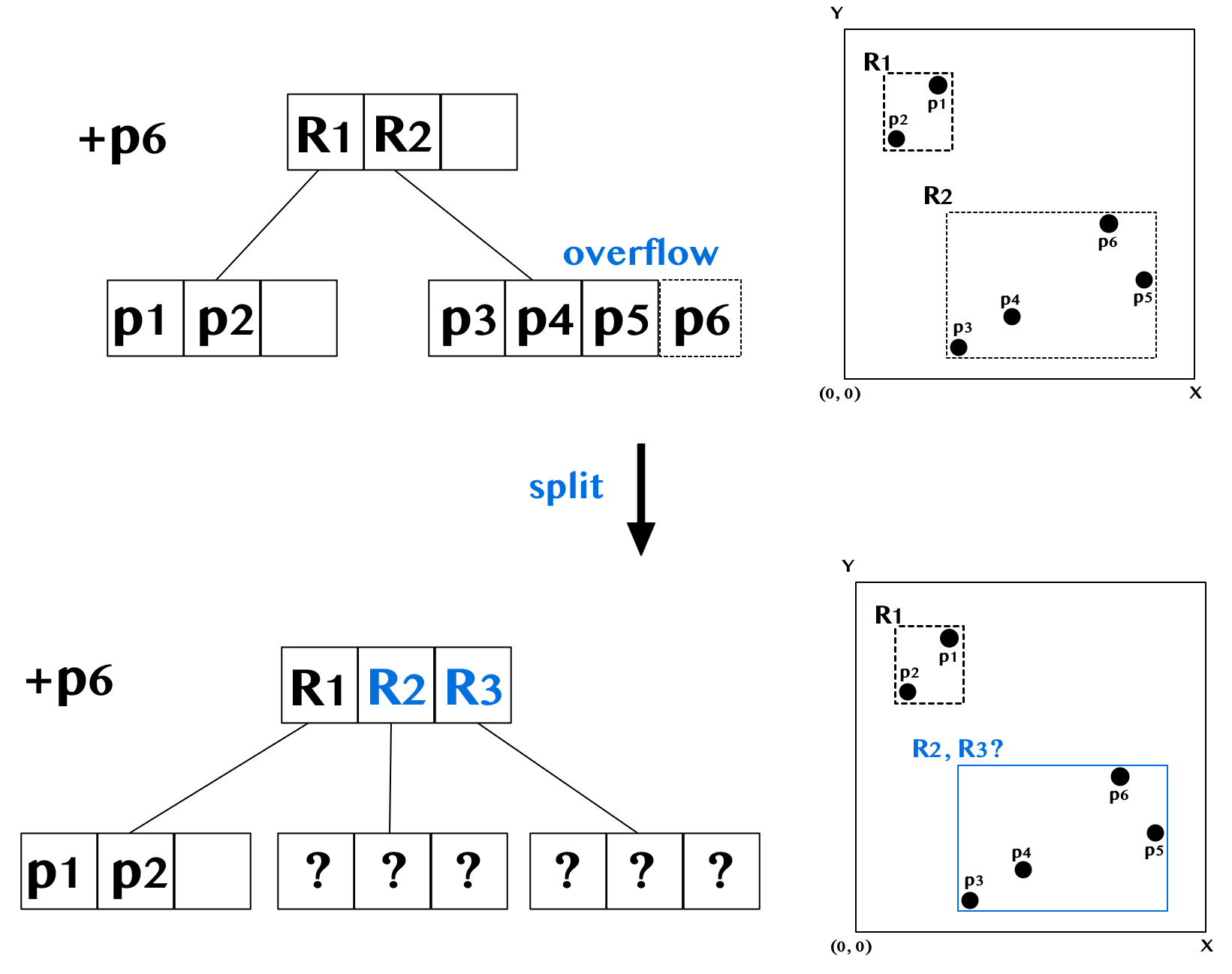}
  \caption{Split case.}
  \label{fig:tree_insertion_full_node}
\end{figure}

The way the entries are distributed will be decisive in order to achieve an optimal distribution of the objects. A good split will imply faster searches, making the most of the R-tree. However, a bad split can lead to an overlapping of the MBRs which may cause redundant searches throughout the nodes. Therefore, the time complexity of a search on an R-tree with no overlapped MBRs is $O(log_{m}(N))$, otherwise is $O(N)$ on the worst case.

Multiple heuristic algorithms have been proposed in order to handle node splits, each one with its cons and pros. One of the most popular heuristics is the quadratic split \cite{guttman}. The strategy of this algorithm consists of separating the two objects that, together, would create the rectangle with the biggest area which in turn means more dead space. Thereafter, the rest of the entries are greedily distributed in one node or the other considering the minor area increment, taking into account the minimum \textit{m} number of entries that a node must have. 
\subsection{K-Nearest Neighbour}

A KNN algorithm consists of, given a reference point, the search of the \textit{k} nearest objects to it. The R-tree data structure allows the execution of this kind of query \cite{Roussopoulos:1995:NNQ:568271.223794}.

The combination of a Branch \& Bound search algorithm together with suitable metrics  \cite{Roussopoulos:1995:NNQ:568271.223794} allows an optimal traversing of the tree. The algorithm prioritises the search through those nodes that are close to the reference point while pruning the farthest ones.
% section about the original structure

% section about the custom implementation
\section{\textit{Ad hoc} implementation}

The particular problem to solve, together with the knowledge of the kind of data to handle, makes possible the implementation of a custom R-tree. This \textit{ad hoc} implementation is strongly based on the original data structure \cite{guttman} but modifying the indexing and search algorithms.

In terms of structure, there is only one significant change from the original description. Each entry from a leaf node contains the hit coordinates. The original R-tree \cite{guttman} stores rectangle objects. In this particular case, points are stored and, therefore, the minimum and maximum value for each axis are the same. Also, leaf nodes contain points, which involves a different way to handle splits on those nodes. The way the R-tree has been represented from section \ref{sec_r_trees} follows such description, with the only difference regarding the implemented version that handles 3D objects (hits) instead of 2D ones.

\subsection{Split algorithm}

About the split algorithm, we have implemented the quadratic split \cite{guttman} as well as a custom split based on a generic pixel tracker geometry. The quadratic split is executed as explained on the original paper \cite{guttman} but changing the way the \textit{PickSeeds} algorithm \cite{guttman} works on leaf nodes. Since the entries for that kind of nodes contains hits instead of rectangles, the two farthest hits are first found and then separated on new nodes.

The custom split algorithm is based on the needs of the CA  reconstruction algorithm. As explained in section \ref{section_pixel_track_reco}, tracks are reconstructed module by module, starting from the back part of the detector. Each module has a unique $Z$ value, implying that all the hits from the same module can be grouped together under the same $Z$. Hence, filtering groups of hits by its $Z$ value and also by closeness among them would considerably prune the number of nodes to be visited each time a KNN search is executed. Further detail of this procedure will be explained in section \ref{label_adhoc_knn}.

Accordingly, every time a split is needed, hits are separated on the two resulting nodes. The hits with the lowest value of \textit{Z} go together in one group while those with the highest value are gathered together on the other. With this procedure, one can obtain an efficient distribution only checking the Z axis. We might be creating more dead space on \textit{XY} in exchange for a reduction of the computation since the time complexity of this custom split is $O(M)$ while the quadratic split \cite{guttman} is $O(M^{2})$.

\subsection{KNN algorithm}
\label{label_adhoc_knn}

\begin{figure}[H]
    \centering
    \includegraphics{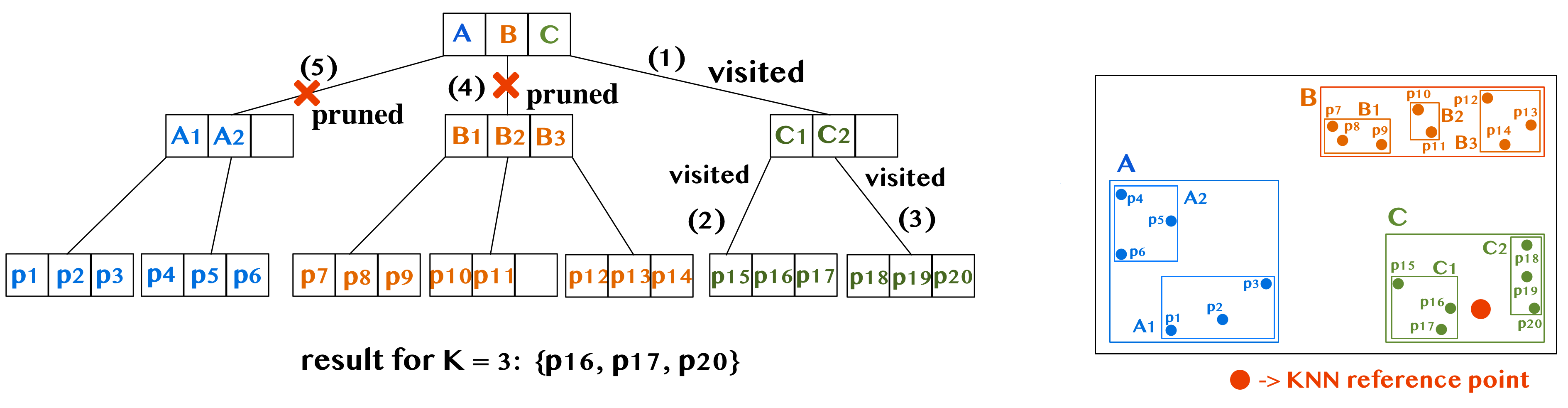}
    \caption{KNN execution example, only 3 nodes visited plus the root.}
    \label{fig:rtree_knn}
\end{figure}

A custom KNN algorithm has been implemented in order to get the $k$ nearest hits of a point in the geometric space of the pixel detector. Since all the searches are executed with the purpose of getting the best candidate hits in specific modules, the reference point for all the KNN calls will belong to a module with the same $Z$ value. This means that the reference  of the R-tree point will be always covered by an MBR as in figure \ref{fig:rtree_knn}, instead of being on a non-covered area.

The previous assumption makes possible the traversing of the R-tree by just checking if the $Z$ values of any entry MBR and the reference point intersect, in order to find nearby hits. Together with a priority queue, we only keep the best $k$ hits based on the distance to the reference point for the $XY$ axis.

The algorithm allows the addition of multiple constraints represented as a dynamic number of user-defined functions, retrieving only the nearest \textit{k} hits that satisfies all of them. A case example is described on section \ref{results_ad_hoc_knn_query} where one constraint function is set to get better candidates. 

\section{Detector model}
\label{sec_detector_model}

\begin{figure}[H]
\centering
\begin{minipage}{.5\textwidth}
  \centering
  \includegraphics[width=.6\textwidth]{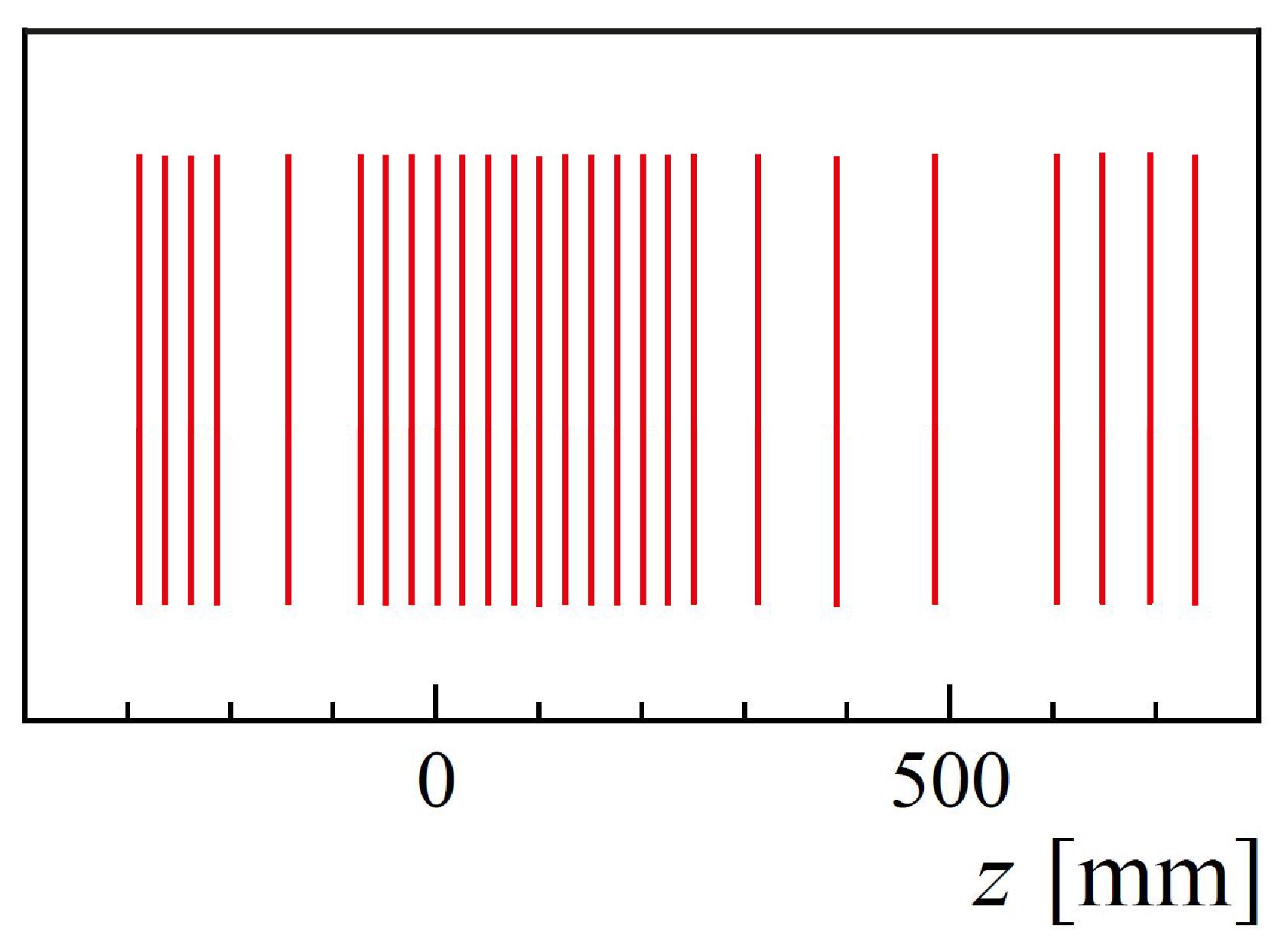}
  \caption{Z position of the sensor modules.}
  \label{fig:zaxis}
\end{minipage}%
\begin{minipage}{.5\textwidth}
  \centering
  \includegraphics[width=.74\textwidth]{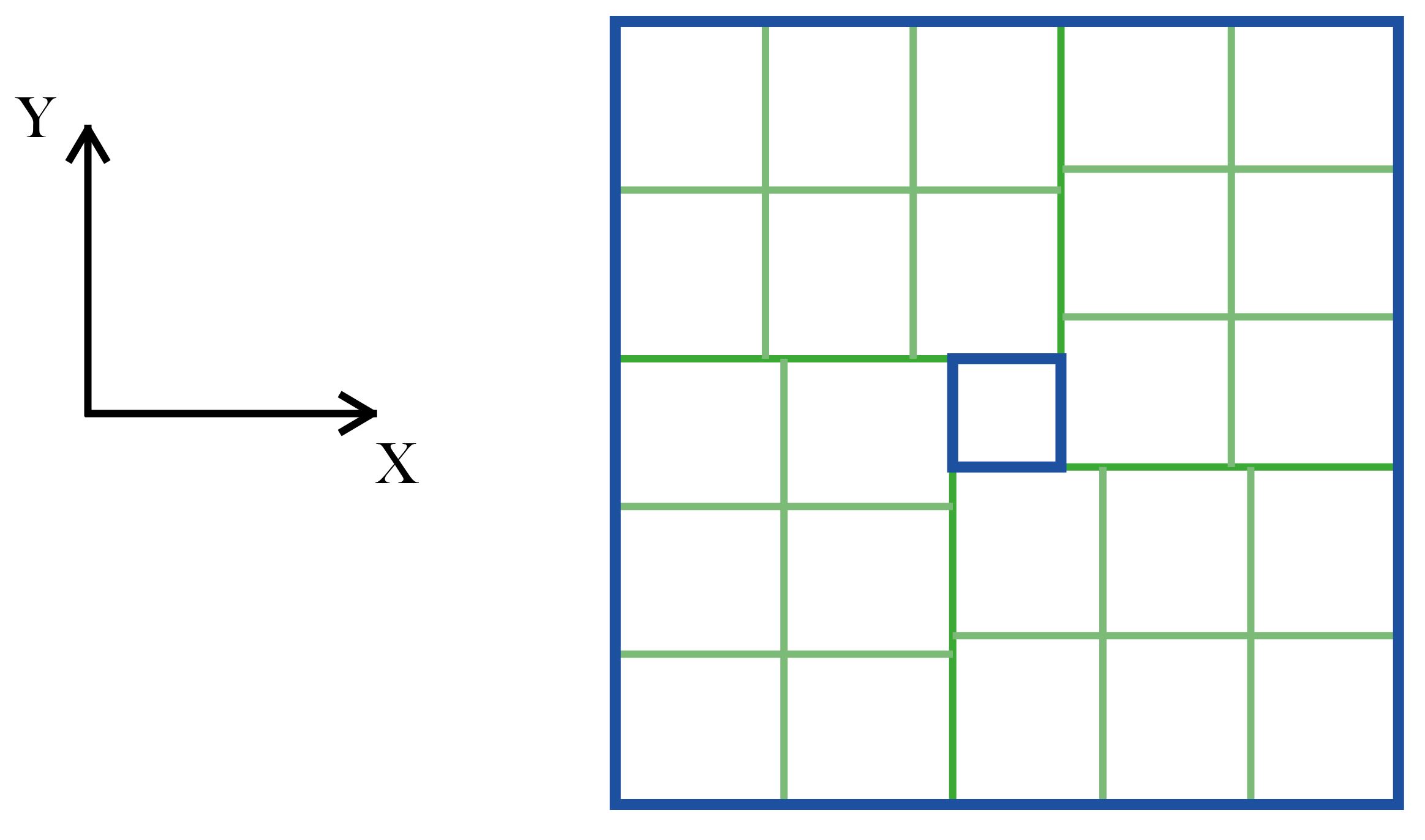}
  \caption{Sensor module geometry.}
  \label{fig:sensor}
\end{minipage}
\end{figure}

In order to simulate the performance of the proposed algorithm a generic pixel detector model is proposed. Highly inspired in the VELO Pixel detector, it consists of a set of $26$ sensor modules placed along the $Z$ axis in specific positions shown in figure \ref{fig:zaxis}. The geometry of each of the modules can be seen in figure \ref{fig:sensor} and consist of a square-shaped plate of dimensions $70.400 mm \times 70.400 mm$ with a squared hole for the beam pipe of dimensions $5.100mm \times 5.100mm$ placed in the $XY$ coordinate origin. All the sensor plate is filled with $55\mu m \times 55 \mu m$ pixels with a detection efficiency set to $1$ for simulation purposes.

\begin{figure}[H]
\centering
\includegraphics[width=.45\textwidth]{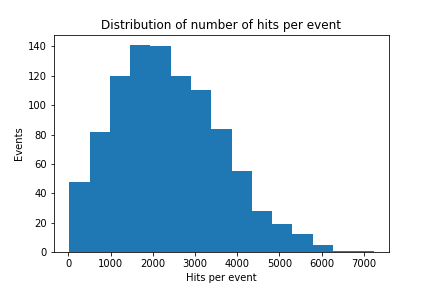}
\vspace*{-1mm}
\caption{Distribution of hits per event in the dataset.}
\label{fig:hitdist}
\end{figure}

The data used to test the efficiency of the algorithm proposed has been extracted from simulations of Monte Carlo particles in the LHCb Upgrade conditions \cite{LHCb-TDR-015} interacting with the geometry proposed above to generate hits. There is a total of $966$ simulated events with a hit distribution per event that can be seen in figure \ref{fig:hitdist}.

For the evaluation of the results obtained there have been used three concepts related to the reconstruction efficiency that will be defined below:

\begin{enumerate}
    \item Reconstruction Efficiency (RE): Defines the number of tracks that fully match with a Monte Carlo track over the total tracks.
    %\begin{equation}
    %    RE = \frac{\#correctly\_reconstructed\_tracks}{\#MC\_tracks}.
    %\end{equation}
    
    \item Clone Fraction (CF): Defines the number of reconstructed tracks that are associated with an already reconstructed track (clone tracks) over the total number of Monte Carlo tracks.
    %\begin{equation}
    %    CF = \frac{\#clone\_tracks}{\#MC\_tracks}.
    %\end{equation}
    
    \item Ghost Fraction (GF): Defines the number of reconstructed tracks that do not belong to any Monte Carlo track (ghost tracks) over the total number of tracks.
    %\begin{equation}
    %    GF = \frac{\#ghost\_tracks}{\#MC\_tracks}.
    %\end{equation}
    
\end{enumerate}

\section{Results}
\label{sec_results}

\begin{figure}[ht]
\centering
\includegraphics[width=.6\textwidth]{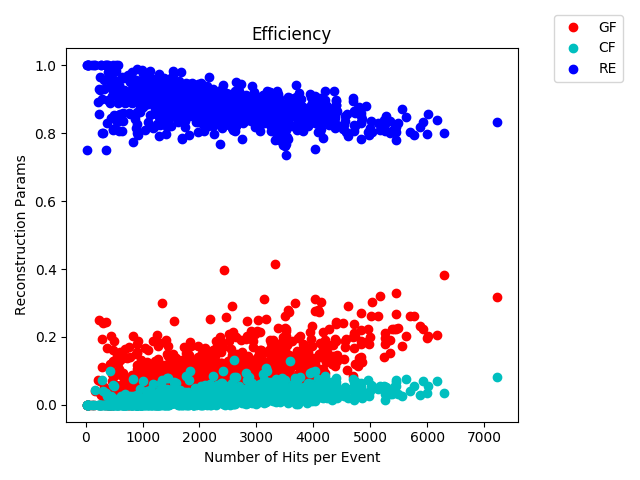}
\vspace*{-1mm}
\caption{Reconstruction parameters as a function of the number of hits per event.}
\label{fig:eff_rtc}
\end{figure}

In order to test the performance of the algorithm implemented, the reconstruction efficiency, the clone fraction and the ghost fraction have been measured through all the $996$ events and can be seen in figure \ref{fig:eff_rtc}.

For comparison purposes, two more implementations of the algorithm proposed have been done. The first one has been implemented maintaining the same execution structure but using a linear data structure for hit indexing. The second one has been implemented using a general purpose R-tree from the Python library C++ Rtree for hit indexing. Table \ref{table:eff} summarises the results obtained as a mean for all the events processed. Also including the results from the \textit{ad hoc} proposed approach.

\begin{table}[h]
\centering
\begin{tabular}{ c|c|c|c } 
  Algorithm & RE & CF & GF \\ 
  \hline
  \hline
 Linear data structure & ($\mu=0.7721,\sigma=0.0724$) & ($\mu=0.0895,\sigma=0.0740$) & ($\mu=0.3232,\sigma=0.1041$) \\
 Generic purpose R-tree  & ($\mu=0.7682,\sigma=0.0675$) & ($\mu=0.0812,\sigma=0.0370$) & ($\mu=0.1496,\sigma=0.0846$) \\
 Ad-hoc R-tree  & ($\mu=0.8815,\sigma=0.0459$) & ($\mu=0.0249,\sigma=0.0216$) & ($\mu=0.1046,\sigma=0.0637$) \\
\end{tabular}
\vspace*{-1mm}
\caption{Reconstruction performances of the three algorithms.}
\label{table:eff}
\end{table}
\label{results_ad_hoc_knn_query}
As it can be seen, there is an improvement in efficiency when using the \textit{ad hoc} implementation of the R-tree in the algorithm. This is because it performs the neighbour search in a more accurate way in the sense that the neighbours returned surely belong only to the module of interest, in the general purpose R-tree, the neighbours returned are chosen only by a proximity criteria without taking into account the sensor module of the hits, which in fact is important for the algorithm as mentioned in previous sections. As a consequence, the custom made R-tree provides to the algorithm better hit candidates to form a track and so, better efficiency is achieved.

\begin{figure}[ht]
\centering
\begin{minipage}{.6\textwidth}
  \centering
  \includegraphics[width=.8\textwidth]{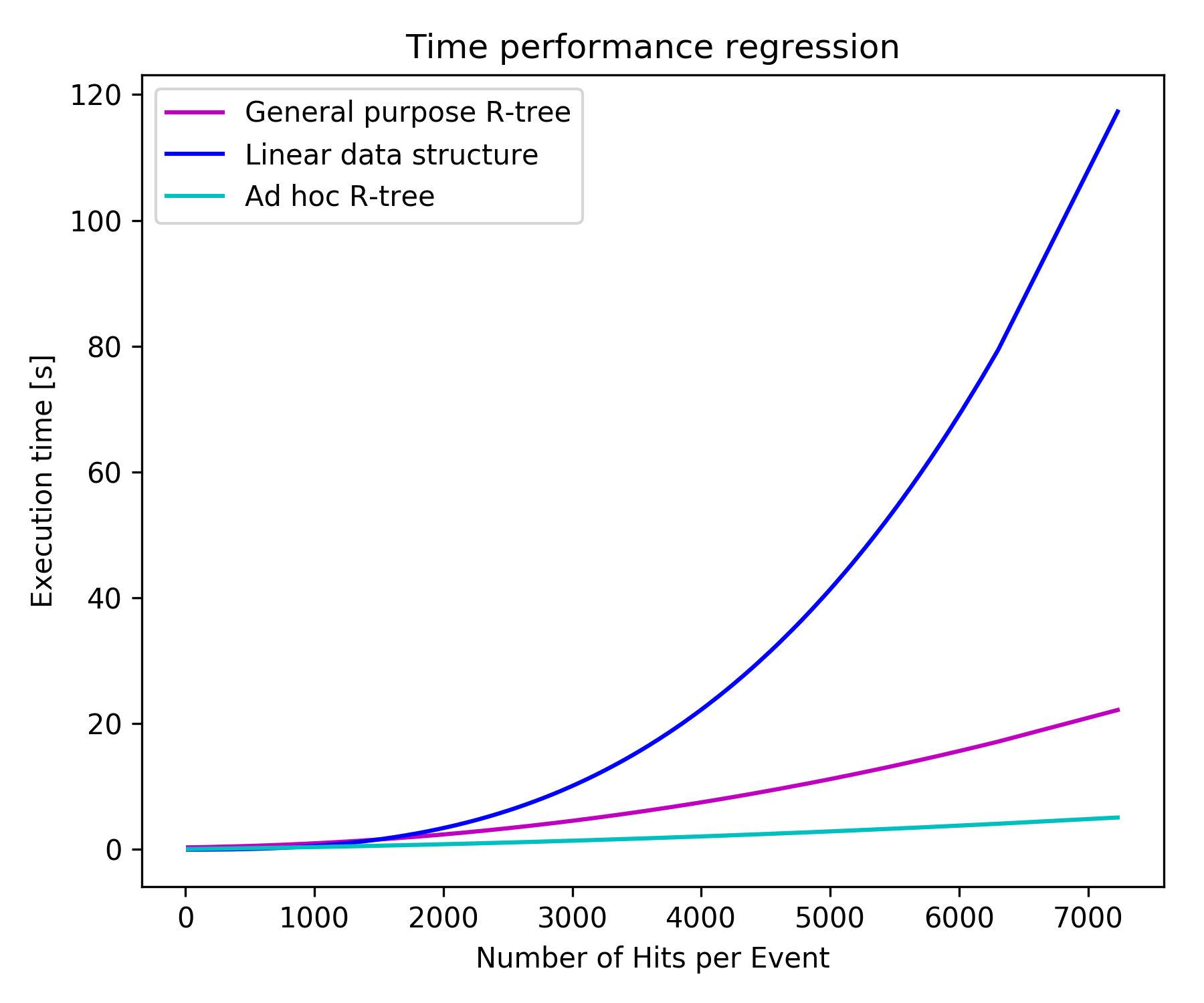}
  \vspace*{-1mm}
  \caption{Reconstruction time regression per hits per event.}
  \label{fig:time}
\end{minipage}%
\begin{minipage}{.4\textwidth}
  \centering
  \includegraphics[width=.9\textwidth]{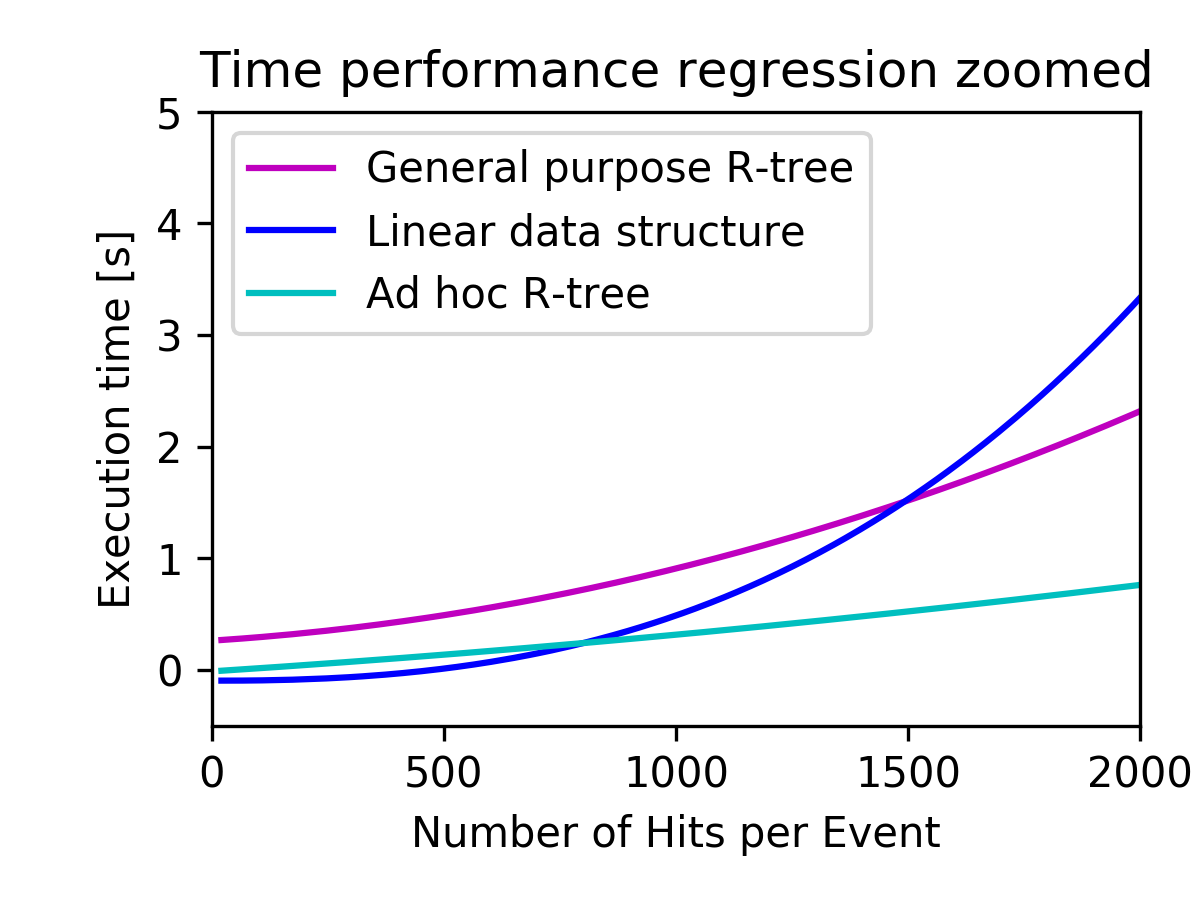}
  \vspace*{-1mm}
  \caption{Zoom of the initial part of the time regression per hits per event.}
  \label{fig:time_zoom}
\end{minipage}
\end{figure}

A study has also been done regarding the computational time of execution for the three algorithms. The results shown in figure \ref{fig:time} plot the regression of the execution time values given as a mean among $100$ different reconstructions of the same event, for each of the $966$ events as a function of the number of hits per event. 

As it can be seen, the curve of the computational time grows much slower in the \textit{ad hoc} algorithm case. Although, in events with less than $750$ hits, the original algorithm is faster than the implementations using the \textit{ad hoc} R-tree, as can be seen in figure \ref{fig:time_zoom}. This can be explained because when working with a small amount of hits, the full combinatorial search of candidates is faster than the cost of the tree searches. But it is also important to know that only $9,1\%$ of the events in the data set have less than $750$ hits.

\begin{figure}[H]
\centering
\includegraphics[width=.6\textwidth]{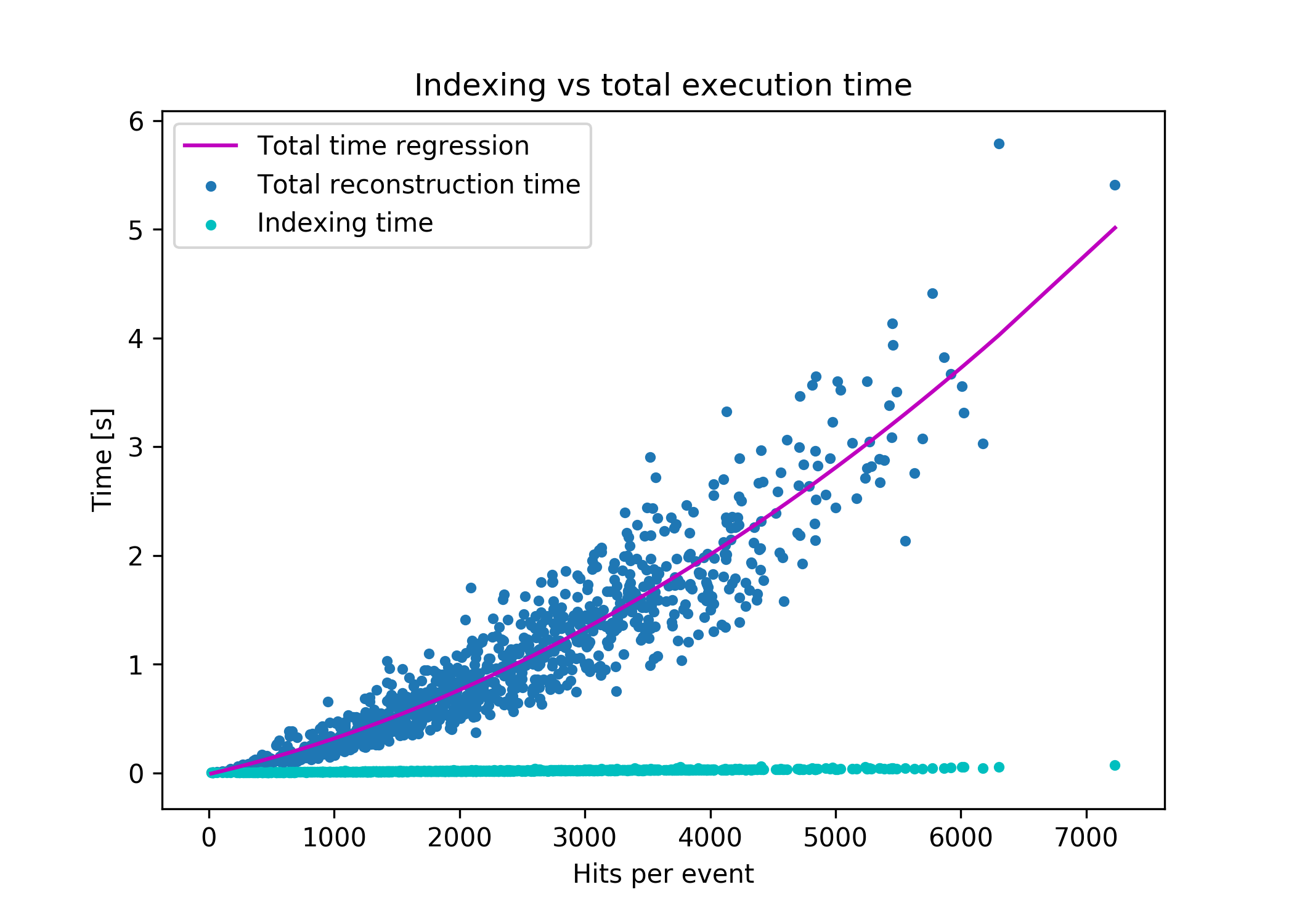}
\vspace*{-1mm}
\caption{Reconstruction and insertion time measures per hits per event using the ad-hoc R-Tree structure.}
\label{fig:time_index}
\end{figure}

Although the computational time curve for full event reconstructions has improved, when one puts it in a proper scale, figure \ref{fig:time_index}, it can be seen that the curve is not linear as it might seem in figure \ref{fig:time}. Instead, it has also a quadratic component that will affect events with an even higher number of hits. Also in the same figure, the insertion time measures are plotted, with this results it can be seen that the added cost to construct the R-tree and fill it with all the event data does not lead to a huge increment on the execution time, instead it has nearly a linear behaviour.
\section{Conclusions}
The above results are a first step in studying the possibilities of advanced data structures, such as R-trees, in providing efficient solutions to cope with high multiplicity events in coming particle detectors. From this toy model, several facts can already be observed. Such structures require preparation, which results in an overhead time. On the other hand, they simplify searches, in the sense of complexity, and deliver faster results. The break-even point is found in our case well below 1000 hits. According to current simulations in the LHC Run3 conditions, this includes most events. The building of the R-tree is controlled by the splitting algorithm, which can be tuned in order to get better efficiency or to reduce ghost tracks. Of course, many open questions remain as future lines of work, such as testing other split algorithms, detector structures and geometries, as well as perusing the knobs of the current tests to better understand how to reduce ghost tracks or deal with missing hits. However, this results point, to us, in a more general direction, into the need of taking care of modern software engineering in order to deal with the coming challenges to be faced by new particle detectors and their needs in terms of data processing.

%%%%%%%%%%%%%%%%%%%%%%%%%%%%%%%%%%%%%%%%%%%%%%%%%%%%%%%%%%%%%%%%%%%%%%%%%%%  

%%  if necessary
\Acknowledgements
This work is supported by the Spanish {\it Ministerio de Econom\'\i a, Industria y Competitividad}, under grant FPA2017-85140-C3-2-P.  

%%%%%%%%%%%%%%%%%%%%%%%%%%%%%%%%%%%%%%%%%%%%%%%%%%%%%%%%%%%%%%%%%%%%%%%%%%%

%\begin{thebibliography}{99}

%%
%%  bibliographic items can be constructed using the LaTeX format in SPIRES:
%%    see    http://www.slac.stanford.edu/spires/hep/latex.html
%%  SPIRES will also supply the CITATION line information; please include it.
%%

%\bibitem{example} 
%  Author 1,
%  "Article title,''
%  Journal {\bf Volume}, Pages (Year)
%  [arXiv:XXXX.YYYY [ZZZZ]].

%\section{References}

\bibliographystyle{ieeetr} % CHECK if is the right one
\bibliography{refs.bib}

\begin{thebibliography}{10}

\bibitem{AtlasPixel}
{ATLAS Collaboration}, ``{Technical Design Report for the ATLAS Inner Tracker
  Pixel Detector},'' Tech. Rep. CERN-LHCC-2017-021. ATLAS-TDR-030, CERN,
  Geneva, Sep 2017.

\bibitem{CMSPixel}
{CMS Collaboration}, ``{CMS Technical Design Report for the Pixel Detector
  Upgrade},'' Tech. Rep. CERN-LHCC-2012-016. CMS-TDR-11, Sep 2012.
\newblock Additional contacts: Jeffrey Spalding, Fermilab,
  Jeffrey.Spalding@cern.ch Didier Contardo, Universite Claude Bernard-Lyon I,
  didier.claude.contardo@cern.ch.

\bibitem{LHCbPixel}
{LHCb Collaboration}, ``{Framework TDR for the LHCb Upgrade: Technical Design
  Report},'' Tech. Rep. CERN-LHCC-2012-007. LHCb-TDR-12, Apr 2012.

\bibitem{cellular_automaton}
I.~Kisel, ``Event reconstruction in the cbm experiment,'' {\em Nuclear
  Instruments and Methods in Physics Research Section A: Accelerators,
  Spectrometers, Detectors and Associated Equipment}, vol.~566, pp.~85--88, 10
  2006.

\bibitem{r_trees_book_theory_and_applications}
Y.~Manolopoulos, A.~Nanopoulos, A.~N. Papadopoulos, and Y.~Theodoridis, {\em
  R-Trees: Theory and Applications}.
\newblock Springer Publishing Company, Incorporated, 2005.

\bibitem{VeloUpgradeTDR}
{LHCb Collaboration}, ``{LHCb VELO Upgrade Technical Design Report},'' Tech.
  Rep. CERN-LHCC-2013-021. LHCB-TDR-013, Nov 2013.

\bibitem{event_reconstruction}
I.~Kisel, ``Event reconstruction in the cbm experiment,'' {\em Nuclear
  Instruments and Methods in Physics Research Section A: Accelerators,
  Spectrometers, Detectors and Associated Equipment}, vol.~566, no.~1, pp.~85
  -- 88, 2006.
\newblock TIME 2005.

\bibitem{guttman}
A.~Guttman, ``R-trees: A dynamic index structure for spatial searching,'' {\em
  SIGMOD Rec.}, vol.~14, pp.~47--57, June 1984.

\bibitem{Roussopoulos:1995:NNQ:568271.223794}
N.~Roussopoulos, S.~Kelley, and F.~Vincent, ``Nearest neighbor queries,'' {\em
  SIGMOD Rec.}, vol.~24, pp.~71--79, May 1995.

\bibitem{LHCb-TDR-015}
{LHCb Collaboration}, ``{LHCb Tracker Upgrade Technical Design Report},'' Tech.
  Rep. CERN-LHCC-2014-001. LHCB-TDR-015, 2 2014.

\end{thebibliography}

%\end{thebibliography}

%%%%%%%%%%%%%%%%%%%%%%%%%%%%%%%%%%%%%%%%%%%%%%%%%%%%%%%%%%%%%%%%%%%%%%%%%%%

\end{document}